\documentclass[apj,graphicx]{emulateapj}
\usepackage{apjfonts}

\usepackage{graphicx}

\newcommand\xmm{{\it XMM-Newton}}
\newcommand\eso{ESO~113$-$G010}
\newcommand\h{1H~0707$-$495}
\newcommand\iras{IRAS~13224$-$3809}

\shortauthors{Cackett et al.}
\shorttitle{Lags in ESO~113-G010}

\begin{document}
\title{A soft X-ray reverberation lag in the AGN ESO~113-G010}

\author{E.~M.~Cackett\altaffilmark{1}}
\author{A.~C.~Fabian\altaffilmark{2}}
\author{A. Zogbhi\altaffilmark{3}}
\author{E.~Kara\altaffilmark{2}}
\author{C.~Reynolds\altaffilmark{3}}
\author{P. Uttley\altaffilmark{4}}

\email{ecackett@wayne.edu}

\affil{\altaffilmark{1}Department of Physics \& Astronomy, Wayne State University, 666 W. Hancock St, Detroit, MI 48201}
\affil{\altaffilmark{2}Institute of Astronomy, University of Cambridge,
Madingley Rd, Cambridge, CB3 0HA, UK}
\affil{\altaffilmark{3} Department of Astronomy, University of Maryland, College Park, MD 20742, USA}
\affil{\altaffilmark{4} Astronomical Institute `Anton Pannekoek', University of Amsterdam, Postbus 94249, 1090 GE Amsterdam, the Netherlands}

\begin{abstract}
Reverberation lags have recently been discovered in a handful of nearby, variable AGN.  Here, we analyze a $\sim$100 ksec archival \xmm\ observation of the highly variable AGN, \eso{} in order to search for lags between hard, 1.5 -- 4.5 keV, and soft, 0.3 -- 0.9 keV, energy X-ray bands.  At the lowest frequencies available in the lightcurve ($\lesssim 1.5\times10^{-4}$ Hz), we find hard lags where the power-law dominated hard band lags the soft band (where the reflection fraction is high).   However, at higher frequencies in the range $(2-3)\times10^{-4}$ Hz we find a soft lag of  $-325\pm89$ s.  The general evolution from hard to soft lags as the frequency increases is similar to other AGN where soft lags have been detected. We interpret this soft lag as due to reverberation from the accretion disk, with the reflection component responding to variability from the X-ray corona.  For a black hole mass of $7\times10^6 \; M_\odot$ this corresponds to a light-crossing time of $\sim$9 $R_g/c$, however, dilution effects mean that the intrinsic lag is likely longer than this.  Based on recent black hole mass-scaling for lag properties, the lag amplitude and frequency are more consistent with a black hole a few times more massive than the best estimates, though flux-dependent effects could easily add scatter this large.

\end{abstract}
\keywords{accretion, accretion disks --- galaxies: active --- galaxies: nuclei --- X-rays: galaxies}

\section{Introduction}

Since the discovery of the first X-ray reverberation lag in the AGN \h\ \citep{fabian09}, X-ray lags between lightcurves from the soft excess and power-law dominated regions of the spectrum (`soft' lags)  have now been observed in over a dozen AGN \citep{zog11_rej1034,emma11,tripathi11,demarco11,demarco12,fabian12} and also in the black hole X-ray binary GX~339$-$4 \citep{uttley11}.  Recently, the first discovery of a broad Fe K lag has now been reported in NGC~4151 \citep{zoghbi12}, with large \xmm\ datasets from \h\ and \iras\ also revealing an Fe K lag \citep{kara12b,kara12a}.  Such lags are naturally predicted in the reflection paradigm where a corona (the power-law component) irradiates the accretion disk, leading to a reflected component consisting of fluorescent emission lines and scattered continuum emission that are broadened and skewed by the dynamical and relativistic effects present in the accretion disk close to the black hole \citep[see][for reviews of reflection]{reynoldsnowak03,miller07}.  The lags can then be ascribed to the light travel time between the direct power-law component and the reflection components \citep[see][for a theoretical discussion on X-ray reverberation]{reynolds99}.

Lags in the opposite sense (`hard' lags where the soft band leads the hard) are also seen on longer timescales.  These hard lags have been seen in both X-ray binaries and AGN  \citep[e.g.][]{miyamoto89, nowak99,papadakis01,arevalo06}, and are thought to arise due to viscous propagation of mass accretion fluctuations in the disk transmitted to the corona \citep{kotov01,arevalouttley06, uttley11}.  Comparing the evolution of the lags with Fourier frequency from all the AGN where they have been detected, a general shape emerges, with hard lags at the lowest frequencies, transitioning to soft lags at higher frequencies, interpreted as viscous propagations dominating on long timescales and reverberation dominating on shorter timescales.  This is not the only proposed model for the lags -- \citet{lancemiller10} suggest an alternative interpretation involving reflection from distant absorbers homogeneously distributed along our line of sight, something that requires a specific geometry.
 While such a model could potentially explain one source, the presence of soft lags in a large number of AGN rules out this model in general \citep[further arguments against this model are given in][]{zog11_1h0707}.

In this manuscript, we present an X-ray timing analysis of \eso\ ($z = 0.0257$), using \xmm observations, which we find to show the same lag-frequency evolution as has been observed in other AGN. \eso\ is a Seyfert 1.8 displaying a strong soft excess and X-ray variability \citep{pietsch98,porquet04,porquet07} which are the two basic components needed for tracing negative reverberation time delays.

\section{Data Analysis}

\eso\ was observed by \xmm\ on 2005/11/10 for a total of 104 ksec, ObsID: 0301890101 \citep[see][for a previous analysis of these data]{porquet07}. We analyzed the \xmm{} data using SAS version 11.0.0 and the most recent calibration files.  Calibrated events files were generated from the Observation Data Files using the epproc and emproc tools for the EPIC/pn and EPIC/MOS detectors, respectively. We searched for background flares in the usual manner, looking at the lightcurves from the entire detector above 10 keV for the MOS, and in the range 10 -- 12 keV for the pn.  We found a significantly elevated background count rate during the beginning and end of the observation, which after we filtered out yielded a continuously sampled data stream (important for timing analysis) of 92ks. 

Spectra and lightcurves were extracted from an 800-px radius circular region, with the
background extracted from nearby source-free region of the same size to the NE of the source for the pn and to the SE for the MOS detectors.  The responses were generated with rmfgen and arfgen, and the spectra grouped to a minimum of 25 counts per bin and a minimum bin width of 1/3 of the FWHM spectral resolution at a given energy using the SAS `specgroup' tool.

\subsection{Spectral analysis}
\begin{figure}
\centering
\includegraphics[angle=270,width=8cm]{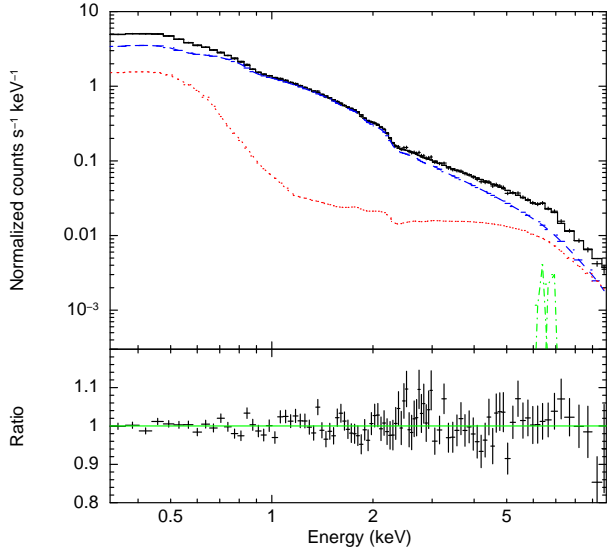}
\caption{\xmm/pn spectrum of \eso\ fitted with a power-law (blue, dashed line), blurred reflection (dotted, red line) and two narrow Gaussian emission lines (green, dot-dashed line).}
\label{fig:spec}
\end{figure}

We fit the pn spectrum in XSPEC v12, in order to determine the energies where the power-law and reflected components are most dominant.  Such a procedure allows us to maximize the sensitivity to finding a lag between these two components by isolating lightcurves where the relative strength of each component  is greatest.  The model consists of an absorbed power-law plus a blurred reflection model using the relconv convolution kernel \citep{dauser10} and reflionx reflection model \citep{rossfabian05}.  We use an unbroken power-law emissivity in relconv.  In addition to this, the data also require an absorption edge at 0.86 keV (rest energy), consistent with an OVIII edge and two narrow emission lines at approximately 6.5 and 6.97 keV \citep[see][for a detailed discussion of the Fe line complex in this object]{porquet07}.   We show the best-fitting model for the pn spectrum in Figure~\ref{fig:spec}, and see figure 7 of \citet{porquet07} for a ratio to a simple power-law.  The spectrum displays a strong soft excess, which can be well-fit by the blurred reflection model (see Table~\ref{tab:specparam} for spectral fitting parameters).  The best-fitting spin parameter indicates a maximally spinning black hole.  While the broad Fe line in this object is not especially constraining on the spin, it is the smooth soft excess that requires a maximally spinning black hole.

\begin{deluxetable}{ll}
\tablecolumns{2}
\tablecaption{Spectral fits to \eso}
\tablehead{Parameter & Value}
\startdata
$N_{\rm H} (10^{20}$ cm$^{-2})$ & 2.74 (fixed) \\
$E_{\rm edge}$ (keV) & $0.86\pm0.01$ \\
$\tau_{\rm max}$ & $0.26\pm0.01$ \\
$E_{\rm line}$ (keV) & $6.52_{-0.05}^{+0.03}$  \\
$\sigma_{\rm line}$ (keV) & 0.0 (fixed) \\
N$_{\rm line}$ &  $(1.7\pm0.5)\times10^{-6}$\\
$E_{\rm line}$ (keV) & 6.97 (fixed)  \\
$\sigma_{\rm line}$ (keV) & 0.0 (fixed) \\
N$_{\rm line}$ &  $(2.4\pm0.5)\times10^{-6}$\\
P.L. index, $\Gamma$ & $2.36\pm0.01$ \\
P.L. norm & $(1.51\pm0.01)\times10^{-3}$ \\
Emissivity index & $8.4^{+0.1}_{-0.9}$ \\
Spin parameter, $a$ & $0.998_{-0.002}$\\
Inclination, $i$ (deg) & $66\pm2$ \\
Fe abundance (Fe/solar) & 1.0 (fixed) \\
Ionization parameter, $\xi$ (erg cm s$^{-1}$) & $1.5_{-0.1}^{+0.8}$ \\
Reflionx normalization & $(8.0\pm1.6)\times10^{-5}$ \\
$\chi_\nu^2$, dof & 1.03, 154
\enddata
\label{tab:specparam}
\tablecomments{We fit the model phabs*zedge*(power-law + zgauss + zgauss + kdblur$\otimes$relionx), and fix $z = 0.0257$ in all relevant model components.}
\end{deluxetable}

Based on the shape of the spectrum, we choose the 0.3 -- 0.9 keV energy range for the soft band lightcurve, where the reflection component is strongest (relative to the power-law), and the 1.5 -- 4.5 keV range for the hard band lightcurve, where the spectrum is dominated by the power-law component. Lightcurves in these energy bands were extracted using evselect and epiclccorr, and 20s binning.

\subsection{Frequency-dependent lags}
We first look for lags between the soft and hard bands as a function of
Fourier-frequency in a similar manner as has been done for all the other sources (see references in section 1).  The lightcurves exhibit significant variability \citep[see][and their figure 1]{porquet07}. 
The lags between the soft and hard bands are determined from the
cross-spectrum following the standard Fourier technique detailed in
\citet{nowak99}. Briefly, we calculate the cross-spectrum for each pair of soft
and hard lightcurves, averaging the cross-spectrum in frequency bins and also
averaging the cross-spectra from multiple detectors.  At a
given frequency, $f$, the argument of the cross-spectrum gives the phase
difference between the Fourier transforms of the two lightcurves.  This phase
difference can then be converted to a time lag by dividing by $2\pi f$.  We
therefore determine time lags in each frequency bin by dividing the average
phase in the bin by the middle-frequency of the logarithmic bin.  We determine
errors in the lags following equation 16 of \citet{nowak99}.

\begin{figure}
\centering
\includegraphics[width=8cm]{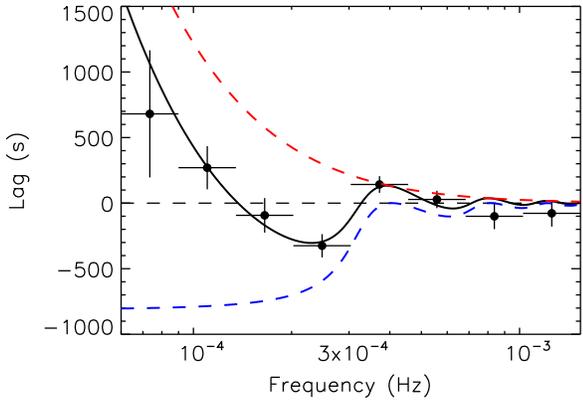} 
\caption{Frequency-dependent lags in \eso{} between the 0.3 -- 0.9~keV and
1.5 -- 4.5~keV lightcurves.  The lag is defined such that a positive lag means
that the hard lightcurve leads the soft.  The solid black line is the best-fitting transfer function, comprised of two components -- a top-hat transfer function (blue dashed line) and power-law transfer function (red dashed line).}
\label{fig:eso113_lag}
\end{figure}

The lags between the soft and hard bands are shown as a function of Fourier frequency in Figure~\ref{fig:eso113_lag}.  The lag is defined such that a positive lag refers to the hard lightcurve lagging the soft (hence `hard' lag), and a negative (or `soft' lag) implies the soft lightcurve lagging the hard.  We find a lag of $-325\pm89$ s in the frequency range $(2-3)\times10^{-4}$ Hz. The evolution of the lags with frequency shows a positive (hard) lag at the lowest frequencies ($\lesssim1.5\times10^{-4}$ Hz) transitioning to a negative (soft) lag at higher frequencies, $(2-3)\times10^{-4}$ Hz, and then becoming a zero lag at the highest frequencies ($\gtrsim4\times10^{-4}$ Hz).  This evolution follows the same general shape as has been seen in the other sources where a soft lag has been detected.  One bin at approximately $4\times10^{-4}$ Hz shows a return to a positive lag.  However, this bin is only 2.2$\sigma$ from zero, and slightly larger frequency bins reduce this significance further.

We model the frequency-dependent lags using two simple transfer functions - a top-hat transfer function  to fit the soft (negative) lags and a power-law transfer function to fit the hard (positive) lags, in a similar fashion to \citet{zog11_1h0707} and \citet{emma11}.  We find that this model readily reproduces the lags.  Our best-fitting model is shown in Fig.~\ref{fig:eso113_lag}, which has the top-hat transfer function extending from 0 to 2390 seconds.  The power-law transfer function is not well constrained as the slope and normalization are degenerate.  However, any combination of those parameters gives the same parameters for the top-hat transfer function.  In lag-frequency space the model for the hard lags is consistent with a power-law with slope of approximately $-1.7$.  This is slightly steeper than the slope of $-1$  seen in other AGN \citep{vaughan03, mchardy04, arevalo08}, though a slope of $-1$ would imply a lag of approximately 100 s at $10^{-3}$ Hz, lying within the 95.4\% confidence interval derived from the data. 

\begin{figure}
\centering
\includegraphics[width=8cm]{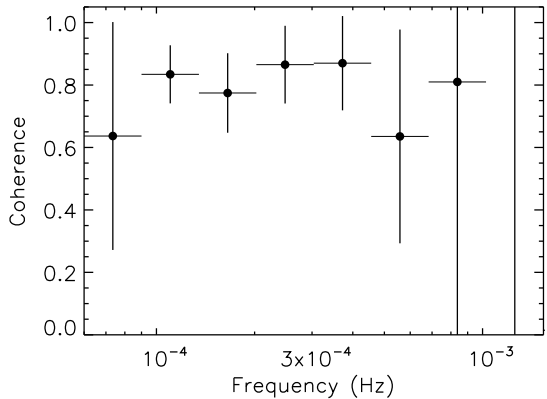} 
\caption{Coherence for \eso{} between the 0.3 -- 0.9~keV and 1.5 -- 4.5~keV
lightcurves.  The lightcurves are coherent to approximately $8\times10^{-4}$
Hz.}
\label{fig:eso113_coherence}
\end{figure}

We check the reliability of the lags by studying the coherence as a function of frequency, following the methodology of \citet{vaughan97}.  The coherence is shown in Fig.~\ref{fig:eso113_coherence}, and remains high over the frequency range where the lags are detected, up to approximately $8\times10^{-4}$ Hz, where the Poisson noise level in the hard band lightcurve is reached.  The high coherence indicates that a high fraction of the lightcurves in each band is correlated and hence one lightcurve can be predicted from the other. 

\subsection{Energy-dependent lags and the covariance spectrum}

\begin{figure}
\centering
\includegraphics[width=8cm]{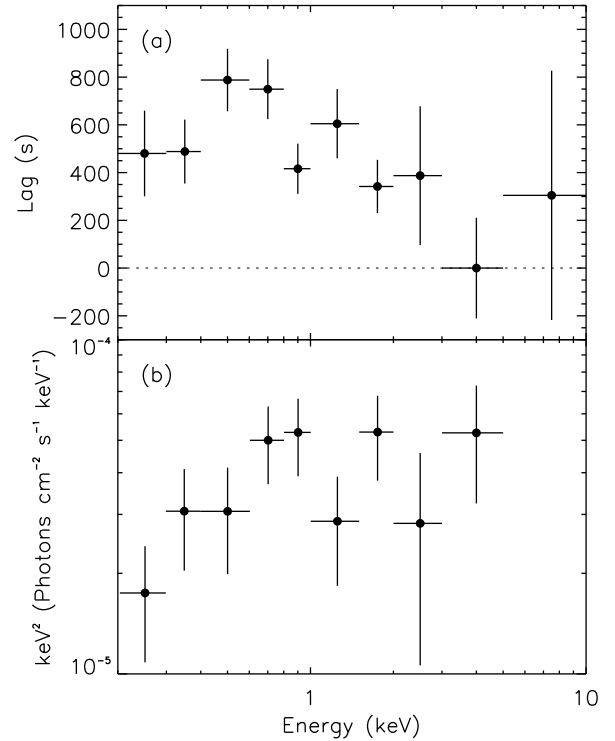} 
\caption{ (a) Energy-dependent lags for \eso{} for the frequency range $(2-3)\times10^{-4}$ Hz, calculated using the pn data only.  The lags are arbitrarily shifted so that the minimum lag is zero.  It is the relative lag between each band that is important. (b) Unfolded covariance spectrum (pn data only) for \eso{} in the same frequency range ($(2-3)\times10^{-4}$ Hz) as the energy-dependent lags in the top panel.}
\label{fig:eso113_lagencovar}
\end{figure}

The energy-dependence of the lags at a given Fourier frequency can help identify the nature and origin of the lags \citep[e.g.][]{zog11_1h0707, kara12b, kara12a}.  Here, we follow a similar procedure to \citet{zog11_1h0707}.  Briefly, we extract lightcurves in narrow energy bands using the pn data only.  Lags are then calculated between each narrow energy band and the reference band.  For the reference band we use the lightcurve from the entire energy range (0.2 -- 10 keV) excluding the current narrow energy band (to avoid correlated Poisson noise).  This ensures a high signal-to-noise ratio in the reference band.  As discussed by \citet{zog11_1h0707} the slightly different reference lightcurve used for  each band leads to an insignificant systematic error in the lag.  In Figure~\ref{fig:eso113_lagencovar}(a) we show the lag spectrum for the frequency range $(2-3)\times10^{-4}$ Hz where the soft lag is detected.  Note that it is the relative lags between the bands that are important (the lags are arbitrarily shifted so that the minimum lag seen is zero).  As was shown in the frequency-dependent lags, it can be seen that the average lag in the soft band is lagging behind the hard band.  It also shows that the lag is largest in the 0.4 -- 0.8 keV range.  If we modify the energy bands for the soft lightcurves slightly to 0.4 -- 0.8 keV this increases the magnitude of the lag observed in the $(2 - 3)\times10^{-4}$ Hz range slightly to $-403\pm83$ s.

It is interesting to note that the general shape of the lag-energy spectrum is quite similar to both \h\ and \iras\ \citep[see][for a comparison of these two objects]{kara12b}, where the lags are seen to peak where the soft excess is strongest, and then drop down to a minimum at 3 -- 4 keV and increase again at the Fe K line.  Obviously, we do not have the signal-to-noise in order to detect a significant increase in the lag above 4 keV (the lag measured in the 5 -- 10 keV range is $305 \pm 522$ s), but, the general shape we see in \eso\ is the same.

We also calculate the covariance spectrum \citep[see][]{wilkinson09} for \eso\ from the pn data, following the methodology of  \citet{uttley11}.  The covariance spectrum shows the strength of correlated variability between a given energy band and the reference band, and hence gives the spectral shape of the correlated variability.  We calculate it in a given frequency range using the frequency-averaged cross spectrum in that range.  The resultant covariance spectrum is proportional to the photon count rate, and so we rebin the pn response to match the energy binning of the lightcurves.  Using the rebinned response, we can fit it in XSPEC.  The covariance spectrum (unfolded using a power-law with index 0 and normalization of 1) is shown in Fig.~\ref{fig:eso113_lagencovar}(b).  A simple power-law with index $\Gamma = 2.1\pm0.2$ fits it well, though given the signal-to-noise (the average fractional uncertainty is 35\%), we cannot rule out more complex models.  Note that the statistics in the 5 -- 10 keV range are too low to calculate the covariance, as is also demonstrated by the large uncertainty for the lag in this energy range.

\section{Discussion}
We have found a lag of  $-325\pm89$ seconds in the variable AGN, \eso\ in the frequency range $(2-3)\times10^{-4}$ Hz.  Moreover, the evolution of the lag over the full frequency range of the lightcurve shows hard lags on the longest timescales changing to soft lags on intermediate timescales and zero lag on the shortest timescales.  This evolutionary trend is the same as in the growing number of AGN where soft lags have been observed.  It suggests common physical mechanisms are present in all these AGN leading to hard and soft lags on different timescales.  The soft lag can be interpreted as due to reverberation from the inner accretion disk, where the lag arises due to light-travel time between the source power-law emission and the accretion disk where the reflected component arises.  Such a model predicts that the soft lag magnitude and frequency where it is observed should both scale with black hole mass, as the characteristic size-scale for the system is simply set by $GM/c^2$.  Recently, \citet{demarco12} observed such scaling with black hole mass using 15 AGN where significant soft lags were detected.

We can compare the soft lag observed in \eso\ with what is expected from the \citet{demarco12} black hole mass scaling. But first, we should consider the black hole mass estimate for \eso.  \citet{porquet07} estimate the black hole mass in \eso\ using the mass-luminosity-timescale relation \citep{mchardy06} to be in the range $(0.4 - 1.0)\times10^7 \; M_\odot$, with the range due to the uncertainty in the bolometric luminosity.  As an additional check on the black hole mass estimate, we use the optical spectrum of \citet{pietsch98} to get a separate estimate.  Optical reverberation mapping has established a scaling relationship between the broad-line region radius and AGN luminosity \citep{kaspi00, bentz_rl_06, bentz_rl_09} which allows single-epoch black hole mass estimates when combined with the broad line width.  While determined using the H$\beta$ emission line, this method has been modified for various other emission lines too, for instance using the stronger H$\alpha$ line \citep{greeneho05,greeneho07}.     \citet{pietsch98} give the H$\alpha$ line FWHM for \eso\ as 2000 km s$^{-1}$.  From their optical spectrum of \eso, we estimate the flux density at 5100\AA\ as $f_\lambda = 1\times10^{-15}$ erg cm$^{-2}$ s$^{-1}$ \AA$^{-1}$.  Using $z=0.0257$ and standard cosmology, we get $\lambda L_\lambda$ (5100\AA) = $7.5\times10^{42}$ erg s$^{-1}$.  We then use the \citet{greeneho05} relationship between $L$(5100\AA) and $L_{H\alpha}$ (their equation 1) along with their updated scaling with $H\alpha$ FWHM from \citet{greeneho07} (their equation A1) to estimate a black hole mass of approximately $7\times10^6 \; M_\odot$, completely consistent with the estimate from \citet{porquet07}.

The lag of  $-325$ s corresponds to 9.4 $R_g/c$ for a mass of $7\times10^6$ M$_\odot$, though dilution of the lag by any hard lag component, and the zero lag component that arises due to reflection and power-law components being present in both bands \citep[see further discussion of dilution effects in][]{zog11_1h0707,kara12b,kara12a}, means that the intrinsic reverberation lag will be larger than this.  Determining the dilution effects is non-trivial given that it depends on the relative strengths of the variable components, however, fitting of the lags with transfer functions, suggests that the intrinsic lag may be closer to 800s (see the blue line in Fig.~\ref{fig:eso113_lag}).  Note, however, an added complication is that the observed lags are not adjusted for gravitational (Shapiro) time delays which would act in the opposite sense to dilution effects.  Comparing with the soft lag magnitude -- black hole mass relation of \citet{demarco12}, the lag lies significantly above the scaling relation and would be more consistent with a black hole mass a factor of a few higher.  \citet{demarco12} also present the scaling of the frequency where the soft lag is observed with black hole mass (the frequency decreases with increasing black hole mass).  The frequency where the soft lag is observed (approximately $2.5\times10^{-4}$ Hz) is slightly lower than expected from the scaling, again consistent with a black hole a few times more massive.  However, it is important to note that scatter is expected in these scaling relations, especially given that one source, \iras\ has a soft lag that is seen to both increase in magnitude and decrease in frequency by a factor of approximately 3 in the high flux state compared to the low flux state \citep{kara12b}.  Thus, flux-dependence could add significant scatter to these lag scaling relations.  Moreover, several of the soft lags detected in \citet{demarco12} occur at the lowest frequencies sampled by the lightcurve and the real minimum may well occur at even lower frequencies not sampled by the current data.  This too can add scatter to the lag scaling relations.

Another cause of scatter in the \citet{demarco12} relations is the black hole spin.  Black hole spin changes the location of the inner disk radius \citep{bardeen72,thorne74}, thus objects with a maximally-spinning Kerr black hole will have shorter lags than objects with a non-spinning (Schwarzschild) black hole (assuming the rest of the geometry is unchanged).  Therefore, non-spinning black holes could be above the black hole mass-lag scaling relation, without requiring a more massive black hole.  However, our spectral fitting of \eso\  imply an inner disk radius consistent with a maximally spinning black hole, suggesting that black hole spin is not the origin of the offset from the scaling relations.

\acknowledgements
ACF thanks the Royal Society for support.  We thank the referee for helpful suggestions that have helped improve the manuscript.

\bibliographystyle{apj}
\bibliography{apj-jour,agn}

\end{document}